%% file: conference_101719.tex
\documentclass[conference]{IEEEtran}
\IEEEoverridecommandlockouts
\usepackage{cite}
\usepackage{amsmath,amssymb,amsfonts}
\usepackage{algorithmic}
\usepackage{graphicx}
\usepackage{textcomp}
\usepackage{xcolor}
\usepackage{amsmath}
\usepackage{amsthm}
\usepackage{multirow}
\usepackage{multicol}
\usepackage{algorithm}
\usepackage{algorithmic}
\usepackage{hyperref}
\usepackage[misc]{ifsym}
\usepackage{bbding}
\usepackage{subfig}
\usepackage{pifont}% http://ctan.org/pkg/pifont
\newcommand{\cmark}{\ding{51}}%
\newcommand{\xmark}{\ding{55}}%
\usepackage{booktabs}
\newcommand{\name}{UniLog}
\def\BibTeX{{\rm B\kern-.05em{\sc i\kern-.025em b}\kern-.08em
    T\kern-.1667em\lower.7ex\hbox{E}\kern-.125emX}}
\begin{document}

\title{UniLog: Deploy One Model and Specialize it for All Log Analysis Tasks}

\author{
 \IEEEauthorblockN{
 Yichen Zhu\IEEEauthorrefmark{3},
 Weibin Meng\IEEEauthorrefmark{4}, 
 Ying Liu\IEEEauthorrefmark{2},
 Shenglin Zhang\IEEEauthorrefmark{5}
 } 
 \IEEEauthorblockN{
 Tao Han\IEEEauthorrefmark{4},
 Shimin Tao\IEEEauthorrefmark{4}, 
 Dan Pei\IEEEauthorrefmark{2}
 }

 \IEEEauthorblockA{\IEEEauthorrefmark{3}University of Toronto}
 \IEEEauthorblockA{\IEEEauthorrefmark{4}Huawei}
 \IEEEauthorblockA{\IEEEauthorrefmark{2}Tsinghua University}
 \IEEEauthorblockA{\IEEEauthorrefmark{5}Nankai University}
  k.zhu@mail.utoronto.ca,
  mengweibin3@huawei.com,
   liuying@cernet.edu.cn,
  zhangsl@nankai.edu.cn\\
 billow.han@huawei.com, 
 taoshimin@huawei.com, 
 peidan@tsinghua.edu.cn

}

\maketitle

\begin{abstract}
Log analysis is vitally important for network reliability, and many log-based tasks are derived to analysis logs, such as anomaly detection, failure prediction, log compression, and log summarization. It is desired to have a unified log analysis framework to simultaneously run all these log analysis tasks on one model to achieve deployment convenience, superior task performance, and low maintenance cost. However, due to severe challenges about log data heterogeneity across devices, pioneer works design specialized algorithms for each task. In this work, we formulate the log analysis as a multi-task learning approach and propose to train a single model that can perform various log analysis tasks. We name this unified log analysis approach as \textit{UniLog}. To effectively build an UniLog model, we propose a log data pretrained transformer to utilize the enormous unlabeled log data, and a corresponding multi-log-tasking finetune strategy for various log analysis tasks. Extensive experiments across seven datasets on four log analysis tasks demonstrate that UniLog achieves remarkable performance.
\end{abstract}

\begin{IEEEkeywords}
Log Analysis, Pretrained Model, Transformer
\end{IEEEkeywords}

\section{Introduction}
\begin{figure}[htbp]
      \begin{minipage}{1.0\linewidth}
      \centering
      \includegraphics[width = 7.5 cm]{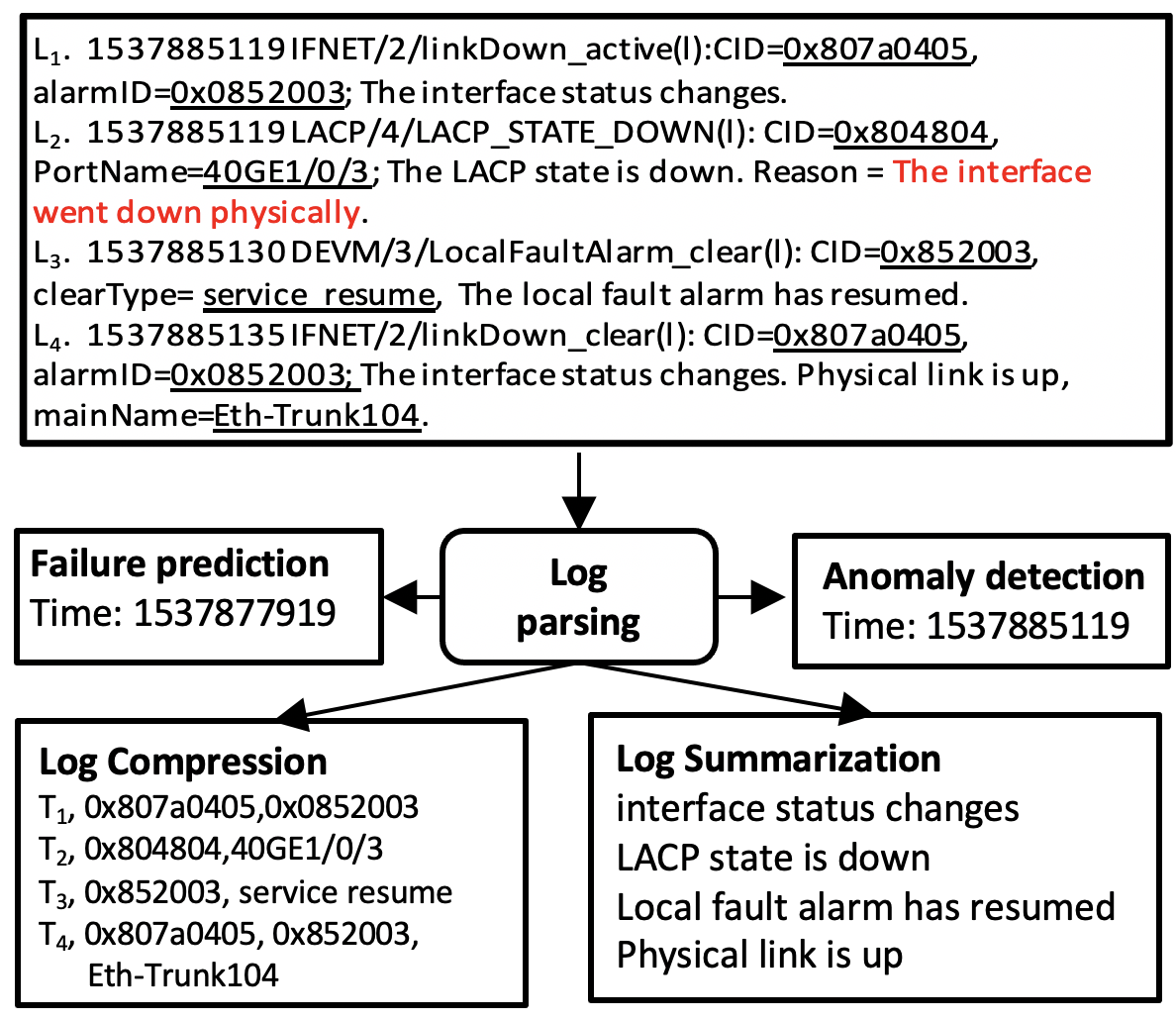}\\
      \end{minipage}
     \vspace{-1 mm}
      \caption{
      Examples of log analysis tasks. Log parsing is a pre-processing step for automated log analysis.
      }\label{fig:logs}
      \vspace{-4 mm}
\end{figure}

Large-scale network services have been increasingly integrated into our daily lives. Most of them are expected to be available on a 24 × 7 basis, and any accidental downtime can lead to irreparable performance and significant revenue loss. For instance, a downtime in Amazon in 2017 led to a loss of around 150 million US dollars~\cite{he2020survey}~\cite{amazondown}. Therefore, network service reliability becomes increasingly important. These network services continuously generate large amounts of logs (see Fig. ~\ref{fig:logs} for examples), which describe a vast range of events and record run-time information. The huge volume of logs makes it impractical to manually inspect logs for key diagnostic information, even with ``search'' or ``grep'' toolkit. Thus there are fast-growing demands to design automatic log analysis frameworks. 

\begin{figure}
      \begin{minipage}{1.0\linewidth}
      \centering
      \includegraphics[height=4cm, width = 7 cm]{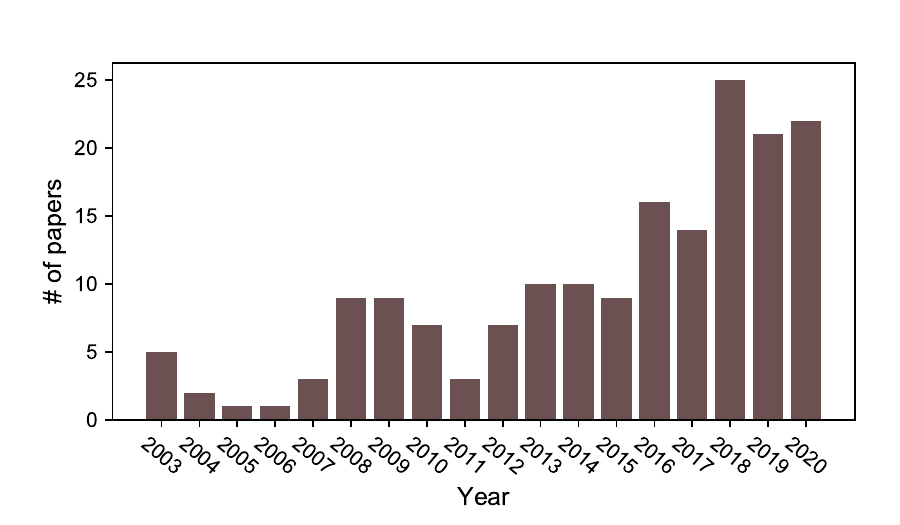}\\
      \end{minipage}
    %  \vspace{-3 mm}
      \caption{The number of log analysis papers published since 2003. We observed that (1) automated log analysis has been continuously and actively investigated in the past two decades, and (2) the number of publications witnesses a sharp rise since 2016.}\label{fig:publication}
    %   \vspace{-3 mm}
\end{figure}

% As shown in Figure~\ref{fig:publication}, 
Many automatic log analysis approaches have been proposed with the success of machine learning and deep learning, as shown in Fig.~\ref{fig:publication}. Existing log analysis approaches can be classified into four tasks~\cite{he2020survey} (see Fig.~\ref{fig:logs}), \textit{i.e.,} anomaly detection aiming to proactively detecting anomalous service behaviors~\cite{meng2019loganomaly}, failure prediction which predicts service failure according to omen patterns~\cite{zhang2018prefix}, log compression trying to save the storage space of massive logs~\cite{logzip}, and log summarization summarizing a series of logs~\cite{meng2020summarizing}.

Despite the tremendous success of the existing log analysis methods, these algorithms only specialize in one type of log analysis task. Nevertheless, real-life deployment constantly desired to do multiple log analysis tasks simultaneously. Imagine a scenario: the maintenance system localizes anomalous logs, and it warns the operators about a potential system failure that might happen very soon. Meanwhile, the system summarizes the unreadable logs into few words to let the operators examine the summarized logs and compress these vast amounts of real-time logs into lossless log data for future usage. More importantly, all these functionalities are performed via a single model, which is easy to debug and upgrade.

Such a model is appealing yet hard to accomplish due to the following challenges. First of all, log data are heterogeneous across domains (i.e., datasets, devices, time.) because different datasets and devices are always filled with different rules to generate logs. Moreover, the style of generated logs from the network system can change from time to time (i.e., due to system upgrade), makes even more challenging to design a good model for log analysis tasks across domains. Secondly, there are many types of log analysis tasks; these tasks are not complemented from the perspective of algorithm design. For example, anomaly detection needs to reconstruct the origin piece of the log.
In contrast, failure prediction needs to predict a specific time where failures might happen. Therefore, current methods design task-specific models for each log analysis task. Unifying these tasks into a single model is non-trivial. 

In this paper, we present a novel approach, named as \name{}, which unified various log analysis tasks under a single framework, enables us to deploy a single model, and specialize it for all log analysis tasks. 
The \name{} has the following merits. First of all, it can generalize to log data from different or even unseen domains. This is achieved by leveraging the Transformer~\cite{vaswani2017attention}, a deep neural architecture that has been rapidly developed in recent years on natural language processing (NLP)~\cite{devlin2018bert}, computer vision~\cite{vit}, and graph data~\cite{graphformer}. We developed a large-scale unlabeled log data pretrain framework to mix up the log data from different domains and use these data to pretrain the Transformer neural networks. Such well designed pretrained Transformer can perform log analysis across different devices and different time periods effectively and consistently.
% second challenges
Secondly, the \name{} is a framework to deploy one model and specialize it for all log analysis tasks. It unifies different log analysis tasks into a single framework. The \name{} would relieve the operators from debugging and upgrading the model to the log analysis tasks themselves. 
% third challenges
Last but not least, switching the \name{} model from one task to another does not involve prior knowledge from the professionals, enabling smooth and automatic workflow for maintaining the network log analysis system.

%Recently, deployment on transformer had shown that such neural network architecture could perform well on target datasets even when the domain of the data varied tremendously. By pretraining on massive unlabeled data from the web, the transformer has dominated natural language processing and the computer vision field. We utilize the transformer architecture and present a pretrain method on large unlabeled log data such that the model can overcome the challenge brought by the log data heterogeneity. We further formulate the log analysis tasks as multi-task learning and present to unified all the log analysis tasks into a single framework. The \name{} has the following merits: 1) It is an end-to-end framework, where the designed training objectives automatically learn the features of log data.  2) It is a task-agnostic unified framework for log analysis, which is applicable on the different deployment scenario 3) It is easy to deploy, maintain, manage, and upgrade. 
Extensive experiments are conducted to verify our proposed \name{} model. We evaluate our method on four log analysis tasks, including log anomaly detection, log failure prediction, log summarization, and log data compression. Our model performs competitively or outperforms state-of-the-art methods on seven public log datasets across these four log analysis datasets. For instance, we outperform PreFix~\cite{zhang2018prefix} by an average of 5.46\% F1 score on failure prediction tasks on Switch log dataset~\cite{zhang2018prefix}. On log anomaly detection, we achieve better F1 scores on both HDFS and BGL datasets than state-of-the-arts. Furthermore, our experiments have shown that the \name{} can generalize on unseen log data, which overcome log data domain adaption and concept drift problem. We believe that \name{} offers an interesting and practical new perspective on designing the model for log analysis tasks.

\input{related_works}
\label{sec:related_works}

\section{Design of \name{}}
\input{methods}

\section{Experiments}
\input{experiments}

\label{sec:experiment}

\section{Conclusion}
\input{conclusion}
\label{sec:conclusion}
%\input{refb.bbl}
% \newpage
\bibliographystyle{unsrt} 
\bibliography{ref.bib}

% \begin{thebibliography}{00}
% \bibitem{b1} G. Eason, B. Noble, and I. N. Sneddon, ``On certain integrals of Lipschitz-Hankel type involving products of Bessel functions,'' Phil. Trans. Roy. Soc. London, vol. A247, pp. 529--551, April 1955.
% \bibitem{b2} J. Clerk Maxwell, A Treatise on Electricity and Magnetism, 3rd ed., vol. 2. Oxford: Clarendon, 1892, pp.68--73.
% \bibitem{b3} I. S. Jacobs and C. P. Bean, ``Fine particles, thin films and exchange anisotropy,'' in Magnetism, vol. III, G. T. Rado and H. Suhl, Eds. New York: Academic, 1963, pp. 271--350.
% \bibitem{b4} K. Elissa, ``Title of paper if known,'' unpublished.
% \bibitem{b5} R. Nicole, ``Title of paper with only first word capitalized,'' J. Name Stand. Abbrev., in press.
% \bibitem{b6} Y. Yorozu, M. Hirano, K. Oka, and Y. Tagawa, ``Electron spectroscopy studies on magneto-optical media and plastic substrate interface,'' IEEE Transl. J. Magn. Japan, vol. 2, pp. 740--741, August 1987 [Digests 9th Annual Conf. Magnetics Japan, p. 301, 1982].
% \bibitem{b7} M. Young, The Technical Writer's Handbook. Mill Valley, CA: University Science, 1989.
% \end{thebibliography}
% \vspace{12pt}

\end{document}

%% file: related_works.tex
\begin{figure}
      \begin{minipage}{1.0\linewidth}
      \centering
      \includegraphics[width = 8.8 cm]{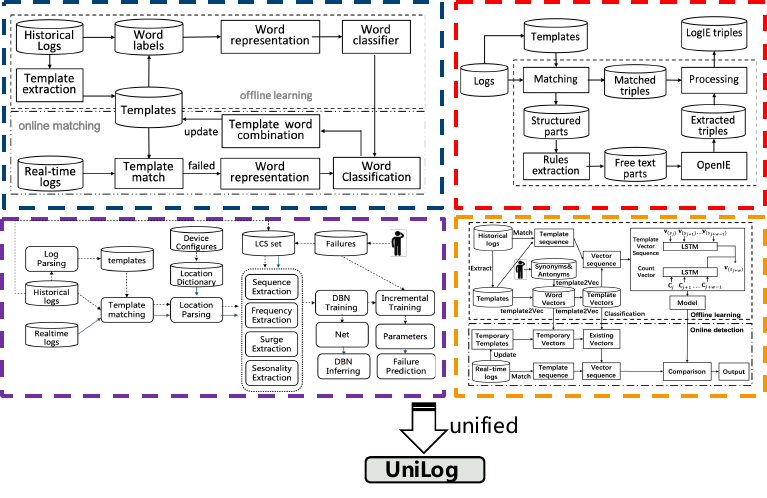}\\
      \end{minipage}
    %  \vspace{-1 mm}
       \caption{An example of existing log analysis frameworks. The top left is a log compression framework. The top right is a log summarization framework. The bottom left is a log failure predication framework. The bottom right are some log anomaly detection algorithms. These methods are designed specifically for a log analysis task. Our approach unified these log analysis tasks in to a single model.}\label{fig:subtasks}
      \vspace{-2 mm}
\end{figure}

\section{Related Work}
\subsection{Transformer}
The Transformer is a type of deep neural network, which have been rapidly developed since Vaswani \textit{et al.}~\cite{vaswani2017attention} and gradually dominate multiple field and tasks, including natural language process (NLP)~\cite{devlin2018bert}, computer vision~\cite{vit, Zhu_2021_ICCV, zhu2021make} and graph~\cite{graphformer}. BERT is the first pretrained model to pretrain the Bi-Directional Transformer on large-scale unlabeled WikiText2~\cite{devlin2018bert} and achieved ground-breaking performance on multiple NLP tasks. Liu \textit{et al.}~\cite{liu2019fine} present to add task-specific classification networks or different losses for each task. The T5~\cite{raffel2019exploring} is another multi-tasking transformer. They convert the NLP tasks to a text-to-text format, which enables to pretrain of the model. Their paradigm presents a new pretrained dataset called C4 and achieves state-of-the-art results in multiple NLP benchmarks. The Transformer has soon been expanded to a huge model that contains billions of parameters~\cite{gpt2} and pretrained on enormous unlabeled data~\cite{gpt3}. Some prior works adopt Transformer or BERT for log anomaly detection. For example, the LogBERT~\cite{guo2021logbert} use Transformer as a log feature extractor. Unlike previous works that only adopt the Transformer as a feature extractor and perform anomaly detection on log data, whereas, we train a single versatile Transformer model that is pretrained on log data and can do multiple log analysis tasks.

\subsection{Log Analysis}
This section briefly reviews prior works on four log analysis tasks, i.e., failure prediction, log anomaly detection,  log compression, and log summarization.
% \subsection{Problem Definition}
% We define the unstructured network logs $\mathbb{X} = \{x_{i}\}$, where $x_i$ is a single network log. In log analysis, we often need a large set of logs to do various tasks $\mathbb{T} = (T_{1}, T_{2}, T_{3}, ..., T_{n})$, where each $T_i$ is a sub task in log analysis. For each task $T_i$, we often want a particular results $y_i = T_i(x_i)$. In this paper, we will discuss four log analysis tasks: log anomaly detection, log summarization, log compression, and log failure prediction. 
% \subsection{Revisit Log Analysis Tasks}
\\
\\
\noindent
\textbf{Log Failure prediction.} 
Network reliability is vital for IT service corporations, and any system failure could cause a significant loss. If the occurrence of system failure can be predicted in advance, then the engineers can take some proactive actions to prevent system collapse and avoid economic loss. The log failure prediction is a supervised binary classification task. Given a log sequence, we need to build a classifier to predict whether the system will corrupt. Existing approaches utilize machine learning algorithms such as XGBoost~\cite{zhang2018prefix} to predict system failure. 
\\
\\
\textbf{Log Anomaly Detection.}
Detecting anomalies in logs is crucial for timely identifying malfunctions of system. Log anomaly detection has been the most active research field for log analysis. The log anomaly detection task is an unsupervised binary task, where each log need to be labeled as normal or abnormal. A conventional way to detect anomalous logs is to use reconstruction method, where log are first encoder into the feature embedding space through a mapping function, and then gradually recover the logs back from the embedding space. By measure the distance with a designed distance metric (\textit{i.e.}, l2 distance) between test data point and normal log feature embedding, we are able to detect the logs that cause system failure. If the calculated distance is larger than a predefined threshold, we will consider it as a anomaly sample. Existing works either use hand-crafted method, recurrent neural network or transformer \cite{deeplog,meng2019loganomaly,guo2021logbert} to extract log feature embedding.
Note that failure prediction is a fundamentally different tasks than log anomaly detection. The anomaly detection aim to locate the anomalous sample in the existing data, whereas failure prediction need to predict the potential anomalous time period that can have anomalous samples in the future.
\\
\\
\noindent
\textbf{Log Compression.} Logs often need to be stored for a long time in practice (\textit{i.e.}, a year) to analyze anomalous problems or root causes. However, storing logs consumes a large amount of storage space and computing resources, which incurs high operational costs. Log compression aims to compress massive log data lossless with the minimum amount of storage. The query time also matters since unzip all logs for a long period, and locating the target log is time-consuming. The compression rate is usually to measure how well an algorithm can compress a certain amount of data to minimize storage space. Log data can be compressed by popular data compression techniques, such as bzip~\cite{bzip}, 7zip~\cite{7zip}, or zip~\cite{zip}. Logparse~\cite{meng2020logparse} is explicitly designed to do log data compression, which has achieved a notable compression rate. This paper integrates the log compression functionality into our framework, improving the compression rate for log data.
\\
\\
\noindent
\textbf{Log Summarization.} 
Logs record detailed run-time information of services. After a failure is detected, diagnosed, or predicted, operators still have to inspect the raw logs to gain a summarized view before taking action. However, manual or rule-based log summarization has become inefficient and ineffective. Log summarization aims to obtain the summarized information of necessary logs for a given log stream. The output is a human-readable sentence. The log summary task is similar to log compression. Nevertheless, the output of the log compression task is not necessary to be readable, as long as its information is lossless. 
In our case, it is necessary to keep the semantics presented in the logs for the output summary to be readable. Meng \textit{et al.} \cite{meng2020summarizing} present a log summarization dataset and designed a log summary framework by using the rule-based algorithm to produce a triplet for each log data. We extend their work to the deep learning approach, which is more general and practical for cross-service log data.

%% file: methods.tex
\begin{figure*}
      \begin{minipage}{1.0\linewidth}
      \centering
      \includegraphics[width = 16 cm]{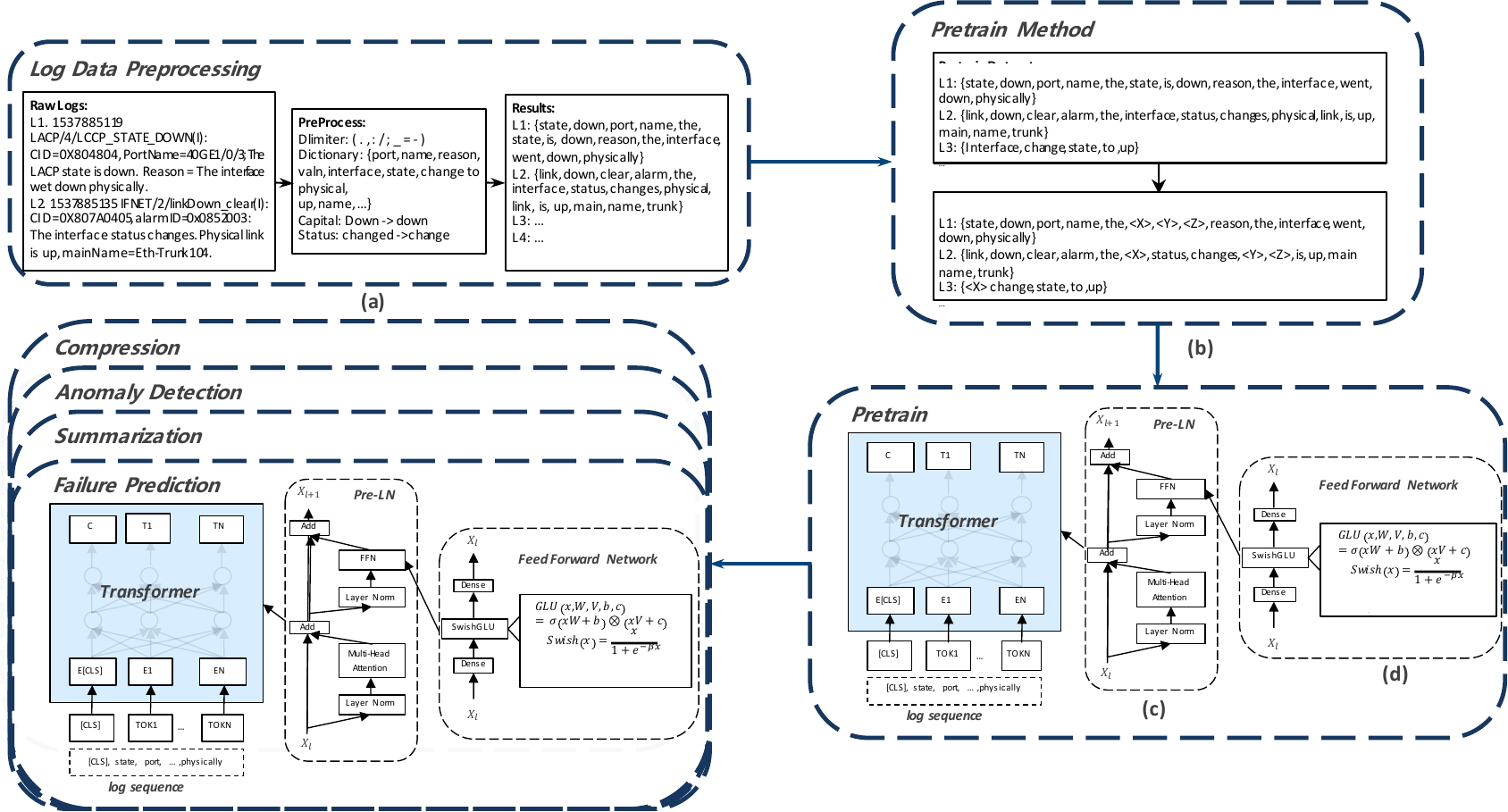}\\
      \end{minipage}
     \vspace{-1 mm}
     \caption{The overview of \name{}. The (a) is the log preprocessing step to prepare the pretrained data. The (b) represents the pretrain method we used in \name{}. The (c) and (d) shows the modification we made in the Transformer neural network architecture.}
      \label{fig:unilog_framework}
      \vspace{-2 mm}
\end{figure*}

In this section, we first introduce two critical components in our \name{} framework: log data pretrained Transformer and multi log analysis tasks finetune strategy. We then discuss some modifications we made in the Transformer neural network architectures.

\subsection{Log Data Pretrained Transformer} 
Pretrained model is prevalent and it dominates many fields in recent years, for example, BERT~\cite{devlin2018bert} in NLP and ImageNet-pretrained backbone in computer vision~\cite{resnet}. When training a neural network that can handle various log analysis tasks, there are two significant challenges. First, log analysis tasks vary, and the desired feature representation of log data is also different for each sub task. Secondly, the log data are heterogeneous across different network devices and different periods. The pretrained Transformer can easily solve these two challenges via 1) the pretrained Transformer can extract \textit{common} log data feature representation from extensive unlabeled data, and these feature representations can be a good initialization point to further finetune to a task-specific feature representation by learning task-specific training objective. 2) the Transformer architecture has been proven that can be effectively scaled up and beneficial when the scale of data increases~\cite{vit}, even when the data are noisy. For the log data, though the meaningful semantic in the log data are scarce, our empirical study demonstrates that it is effective to train the Transformer neural network on large-scale log data.

The pretrain steps are simple. Similar to the natural language, the order feature in the log sequence is important. For example, a system log cannot print ``change state to down'' before the message of ``change state to up'', unless there are anomalous events. Because the pretrain stage do not involved in labels, we choose a self-supervised training objective as the first step to train a unsupervised transformer on log data, which enable the model to understand the importance of log order. A common choice of the training objective is to use BERT-style objective, which randomly sample 15\% of tokens in the input log sequence and then drop them. These dropped token are replaced by a sentinel token and assigned a special token ID to it. The token ID is a special token added to the vocabulary and does not correspond to any wordpiece. The unsupervised training objective thus can be formulated as predicting the dropped out tokens. The output should be the original log sequence.

The natural language is normally considered as structural text: the contexts between consecutive sentences are meaningful. Whereas the log data are unstructured, typically there is no correlation between consecutive log sequences. What matters to the log data is the consecutive words and sentences from long range. Therefore, we proposed a novel pretrain objective based on the particular property of log data. Specifically,
we mask the tokens in the same as we did in the BERT-style pretext task, except that we only mask the consecutive words. We choose up to three consecutive tokens, where the token with largest random probability is chosen. We then mask subsequent two or three tokens. If the chosen token is the last token in the sequence, then we mask the previous two tokens. The output is also the original log sequence. An example of the pretrain objectives can be found in Table~\ref{tbl:pretrain_example}. Secondly, we increase the number of input token to increase the receptive field of log data. As such, the Transformer is able to train with a large chunk of log sequence, and potential learn the long range dependency between two log sequences.
\noindent
\\
\\
\textbf{Log Data Preprocessing.}
The log preprocessing step is briefly shown in Figure~\ref{fig:unilog_framework}(a). We define the word in a log is a token. Tokenization is a crucial step to build a successful log analysis model. However, the structure of the log sequence is quite different from the sentence in natural language. The top part of Figure~\ref{fig:logs} shows few examples of log sequence. Basically, the log sequence contains many meaningless separators and words. Therefore we first use a set of special delimiters (.,:/;\_=-"). These delimiters can primarily separate the log. However, this is insufficient since many words in the logs are not natural language texts, such as ``LocalFaultAlarm\_clear''. After using the delimiter, we still get ``LocalFaultAlarm''. We then use wordninja, a text tokenizer based on English Wikipedia unigram frequencies, to separate meaningful text in a log sequence. In the end, we can get a set of tokens $\{tok_1, tok_2, \dots, tok_n\}$, where each token represents a word. For instance, the word ``LocalFaultAlarm\_clear'' is separated into ``Local'', ``Fault'', ``alarm'', ``clear'' by the first two steps. Additionally, we need to change the capital word to lower case and set a word with a different status, \textit{i.e.,} change and changed, as the same token. 
\\
\\
\noindent
\begin{table*}
\caption{Description for the datasets.}
	\centering
	\begin{tabular}{c|c|cc|cccc}
	\toprule
	\textbf{Datasets}  & \textbf{Description}  & \textbf{\# of logs}  & \textbf{\# of anomalies/failures}  &
	\textbf{A.D.} &
	\textbf{Comp.} & 
	\textbf{Summ.} & \textbf{Fail.}\\
	\midrule
	HDFS   & Hadoop distributed system log & 11,175,629 & 16,838 anomalies  & \cmark  & \cmark  & \cmark  & \xmark\\
	BGL   & Blue Gene/L supercomputer log & 4,747,963 & 348,460 anomalies & \cmark  & \cmark & \cmark  &\xmark \\
	HPC   & High performance cluster log  & 433,489 & - &  \xmark  & \cmark & \cmark   & \xmark\\
	Proxifier  & Proxifier software log & 21,329 & -  & \xmark & \cmark & \cmark  & \xmark\\
	Hadoop  & Hadoop MapReduce job & 394,308  & - & \xmark  & \cmark  & \xmark  & \xmark\\
	ZooKeeper  & ZooKeeper service & 74,380 & - & \xmark & \cmark & \xmark  & \xmark\\
	Switch & Switch hardware failures & 29,174,680 & 2,204 failures & \xmark & \xmark & \xmark & \cmark \\
	\bottomrule
	\end{tabular}

	\label{tab:datasets}
\end{table*}
\textbf{Input and Output.} In order to train a single model on the diverse set of log analysis tasks, we reform the input and output format. We cast all of the tasks we consider into a log-to-log format. Specifically, the model is fed by log sequence and then returns some output logs. This framework provides a consistent training objective to pretrain the \name{} on unlabeled log data and then finetune to fit a specific task. In this way, the \name{} will focus on extracting common feature representations in the log data. Then the model can switch to learn task-specific features by finetuning. To specify in which task the model should be finetuned or performed, we add a task-specific prefix to the input log data sequence before feeding it to our model. For example, we add ``\textless anomaly\textgreater''  before the log sequence if we desire to perform the log anomaly task. The model will then give the output to determine whether a log sequence is normal or anomalous. This holds the same for all tasks, thus allowing the model to handle multiple tasks. The list of a prefix we used is \textless anomaly\textgreater, \textless summarization\textgreater, \textless compression\textgreater, and \textless failure\textgreater, corresponding to log anomaly detection, log summarization, log compression, and failure prediction. 

\subsection{Log Analysis Task-Specific Finetune}
\textbf{Task-Specific Finetune Strategy.} The pretrained log data-based Transformer provides us the common feature representations that universally exist on log data. In order to run specific log analysis tasks such as anomaly detection, we need to finetune our model based on the task training objective so that the model can learn task-specific feature representations. As the log analysis contains multiple sub-tasks, the pretrained model needs to conduct multi-log-tasking. However, perform multi-tasks on a single network can lead to suboptimal results, especially on low-resource tasks \cite{peters2019tune} due to the co-adaptation effect. \name{} aims to process a sequence of logs and then return the output for every log analysis task. The previous method~\cite{T5} gradually unfreeze the parameters in the model to achieve this goal; however, it is sensitive to the size of the training dataset and the total number of training iterations~\cite{raffel2019exploring}. We relax the multi-task finetune framework and introduce a straightforward checkpoint-based method. Specifically, we first record the checkpoint after the model finishes pretraining. Then we load this checkpoint every time a new task needs to be finetuned. During inference, we load the checkpoint for each finetuned task model and return the outputs. Therefore the engineers can simply deploy \name{}, which is a single, unified model, and perform multiple log analysis tasks at the same time. Since there is only one model for deployment, the maintenance cost can be very cheap, and model upgrading can be done under a single framework. The UniLog improves the system reliability, enables the engineers to free their hands on debugging the deployed model, and focuses on system maintenance itself.

Additionally, we introduce extra task-aware heads to finetune on the sub-tasks. Each task-aware head is attached with two consecutive multilayer perceptrons after the log feature embedding layers. A task-aware head is used to disentangle the feature representations that are learned between different log analysis tasks. For instance, there are four log analysis tasks in our framework; then, we will have four task-aware heads, where each head corresponding to a sub-task and return the task-specific results. We can jointly optimize the task-aware head with our pretrained model during finetuning to achieve better model performance. 

For the log failure prediction, we use the softmax operation to return the probability of a network system failure shortly. We use the cross-entropy loss as our training objective. The log anomaly detection use l2 norm loss as the pretrain objectives. This heuristic way is very effective since the pretrain objective loss can represent how hard it is to reconstruct an input log sequence. Similar to the other reconstruction-based anomaly detection method, we set a predefined threshold as 1e-3 and assign the log sequence with validation loss higher than this threshold as the anomalous sample. We use the cross-entropy loss to train the log summarization task to output the summarized log given a sequence of the log. The log compression is more complex than the other tasks, thus we discuss it in a separate section.
\\
\\
\noindent
\textbf{Log Compression.} Besides the transformer network that extracts the log feature embedding, the log compression requires an entropy coding scheme. We choose the arithmetic coding scheme \cite{witten1987arithmetic} for simplicity. Specifically, given a log sequence $s = {s_0, s_1, \dots, s_w}$of length $w$ number of tokens, we first use the predictor $\mathcal{T}$ to encode this stream of tokens and return the probability $P(s) = {P(s_0), P(s_1), \dots, P(s_w)}$ for each token, where $P(s) = P(s_w|s_{w-1}, \dots, s_{0})$ is a token conditional probability depends on the past tokens. In our context, the transformer we describe above can be considered as a predictor. Then an entropy coding scheme is used to store the token into a lossless code of bit-length $log_2P(s_{t}|s_{t-1}, \dots, s_{0})$ based on the token probability given by the predictor. We iteratively compress each token until all tokens are compressed. The decompression stage is similar to the compression stage, except that the entropy coding scheme generates the preceding tokens. Thus the transformer network does not involve in decompression, making the decompression efficient.

\subsection{Details of the Transformer Architecture}
\noindent
\textbf{The Transformer Model.} \label{mtd:model}
After a log sequence is preprocessed into a sequence of log tokens, it is mapped to a sequence of embeddings and then passed into the encoder. The encoder is a stack of transformer blocks, and each block comprises a multi-head self-attention layer~\cite{vaswani2017attention} followed by a shallow feed-forward network (FFN). We apply layer normalization to the input of each self-attention layer and FFN. We also apply Dropout~\cite{srivastava2014dropout} within the FFN, on the residual path, on the attention weights, and at the input and output of the entire stack. The decoder structure is similar to the encoder, except there is a standard attention layer after each self-attention layer. The self-attention layers in the decoder use a form of causal self-attention. Unlike self-attention in the encoder, which allows the model to attend to past outputs and future outputs, causal self-attention only allows the model to attend to past outputs. The decoder's outputs are forward to a fully connected layer with softmax, whose weights are shared with the input embedding matrix. We apply relative position embedding and share this position embedding across all layers in our model. More detailed explanations of the Transformer neural network architectures can be found in ~\cite{devlin2018bert}. In the next section, we introduce two modification on Transformer architecture to improve the model performance for log analysis tasks.
\begin{table*}
\caption{Examples of different pretrained method. The BERT-style objective includes a corruption term, where some token are replace by a random token, in our example, the word link in the original text is replace by word \textit{status}. The original text denote that the model predicts the original input based on the reconstruction objective.}
\centering
% \small
% \renewcommand\tabcolsep{4.5 pt} % 调整表格列间的宽度
\begin{tabular}{ccc}
\toprule
Method & Input &  Target\\
\midrule
Prefix Language Modeling  & The interface status changes  & Physical link is up \\
BERT-style \cite{devlin2018bert}  &  The \textless M\textgreater status \textless M\textgreater Physical \textit{status} is  up  & original text \\
\midrule
\textbf{\name{}} &  The \textless M\textgreater \textless M\textgreater \textless M\textgreater Physical status is  up  & original text \\
\bottomrule
\end{tabular}
\label{tbl:pretrain_example}
\end{table*}
\begin{figure}

      \begin{minipage}{1.0\linewidth}
      \centering
      \includegraphics[height=3.7cm, width = 4.6 cm]{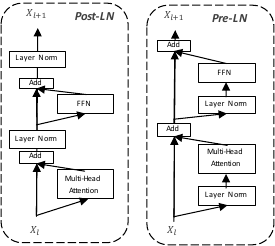}\\
      \end{minipage}
     \vspace{-1 mm}
     \caption{The left is Post-LN and the right is Pre-LN structure which we used in our paper.}
      \label{fig:preln}
      \vspace{-6 mm}
\end{figure}
\\
\\
\noindent
\textbf{Pre-LN Structure.} Layer normalization (LN)~\cite{ba2016layer} is an essential building block of the transformer, and Post-LN architecture is a common choice. However, since we utilize long-range sequences to learn the feature representation of log data, the parameter gradients, which we used to update the model parameters, can be extremely high and cause gradient overflow. Previous study~\cite{xiong2020layer} on Pre-LN structure have shown to possess well-behaved properties on gradients at initialization, thus alleviate the large gradient during training. The comparison of Post-LN and Pre-LN structure are shown in Figure~\ref{fig:preln}. In short, Post-LN places the LN on the main path so that the gradient of the parameters near the output layer are large~\cite{xiong2020layer}. Whereas the Pre-LN place on the residual path. An empirical study is performed in 
in the Fig.~\ref{fig:preln}.
% The place of Layer normalization \cite{ba2016layer} is important to the optimization in Transformer.The comparison of Post-LN and Pre-LN structure are shown in Figure~\ref{fig:preln}. We empirically study the importance of Post-LN and Pre-LN in Table~\ref{tbl:preln}. The position of layer normalization is surprisingly critical in the log analysis task. In the log summarization task, the Pre-LN achieve 1.7\% higher F1 score than the Post-LN. In the log anomaly detection task, the Pre-LN achieve 0.05\% higher F2 score than the Post-LN. These results show the necessity of using Pre-LN in our model.
\\
\\
\textbf{LogAct.} Feed-forward network (FFN) is another fundamental component of the Transformer. It is comprised of few stacks of multilayer perceptron followed by non-linear activation. The non-linear activation controls how much expressiveness~\cite{mellor2021neural} can a neural network represent. We introduce a novel non-linear activation function, named as \textit{LogAct}, to enhance the expressive power of the Transformer, which is more suitable to deal with long-range, unstructured log data for log analysis tasks.

Inspired by \cite{dauphin2017language} and \cite{ramachandran2017searching}, our LogAct combines the GLU activation function \cite{dauphin2017language} and Swish activation function \cite{ramachandran2017searching}. Specifically, the GLU unit can be formulated as:
\begin{equation}
    GLU(x, W, V, b, c) = \sigma(xW + b) \otimes (xV + c)
\end{equation}
where $x$ is the input, $W$ and $V$ are weight matrix, $b$ and $c$ are bias. The Swish activation function can be formulated as:
\begin{equation}
    Swish(x) = \frac{x}{1 + e^{-\beta x}}
\end{equation}
where the $\beta$ is a learnable parameters. In our paper, we combine the Swish function and GLU function as:
\begin{equation}
    LogAct(x, W, V, b, c) = \frac{(xW + b)}{1 + e^{-\beta (xW + b)}} \otimes (xV + c)
\end{equation}
In our experiment, we replace very ReLU layer in the FFN with LogAct. We conduct an ablation study on the effectiveness of LogAct in Table~\ref{tbl:relu}.

%% file: experiments.tex
\begin{table}
\caption{Comparison of model performance for failure prediction task.}
\centering
% \small
% \renewcommand\tabcolsep{4.5 pt} % 调整表格列间的宽度
\begin{tabular}{c|lccc}
\toprule
Switch &   Method   &  Precision   & Recall  & F1 \\
\midrule
\multirow{4}{*}{M1} & HSMM  &  32.27\% & 95.3\% & 48.21\% \\
& SKSVM  & 8.25\%  & 76.09\%  & 14.89\% \\
& PreFix  &  \textbf{87.35\%} &  74.36\%  & 80.33\% \\
& \textbf{\name{}}  &  84.25\% &  \textbf{79.60\%}   &   \textbf{81.86\%}  \\
\midrule
\multirow{4}{*}{M2} & HSMM  &  0.28\% & 60.58\% & 0.56\% \\
& SKSVM  &  4.47\%  & 8.72\%  & 5.91\% \\
& PreFix  &  59.79\% &  58.59\%  & 59.19\% \\
& \textbf{\name{}}  &  \textbf{65.73\%}  & \textbf{63.24\%}  & \textbf{64.46\%}   \\
\midrule
\multirow{4}{*}{M3} & HSMM  &  26.32\% & 11.11\% & 15.63\% \\
& SKSVM  & 0.79\%  & 91.91\%  & 1.58\% \\
& PreFix  &  84.00\% &  52.50\%  & 64.61\% \\
& \textbf{\name{}}  &  \textbf{87.90\%} & \textbf{64.58\%}  &  \textbf{74.46\%}  \\
\bottomrule
\end{tabular}
\label{tbl:fail}
\vspace{-4 mm}
\end{table}

In this section, we evaluate our methods on four log analysis tasks: log anomaly detection, log summarization, log compression, and log failure prediction. We briefly summarized the datasets in Table~\ref{tab:datasets}. Note that for some tasks, the dataset may not have labels for evaluation; thus, we omit those experiments (denotes as cross-mark in the table). The details baseline and results will be introduced in the following sections for each task, respectively. Through our experiment section, we sometime use abbreviation for simplicity, i.e., "A.D." to denote anomaly detection, "Comp." to represent log compression, and "Summ." for log summarization.

\begin{figure}[t]
      \begin{minipage}{1.0\linewidth}
      \centering
      \includegraphics[height=3cm, width = 6.5 cm]{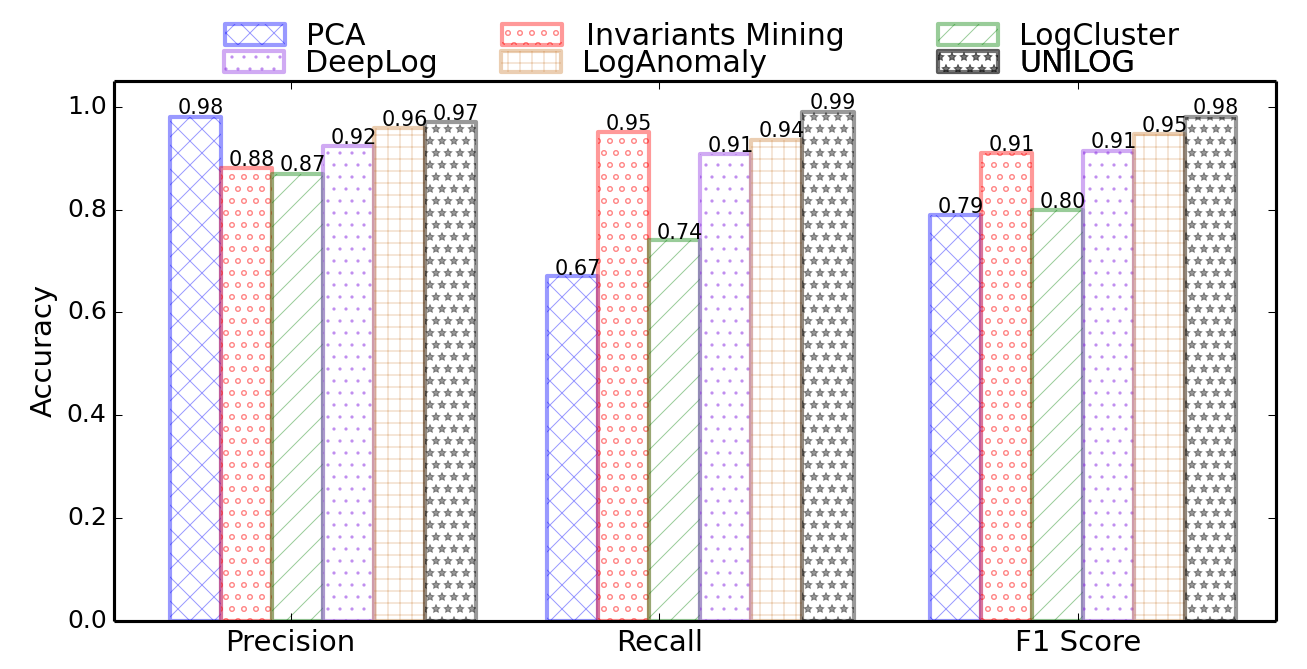}\\
      \end{minipage}
     \vspace{-3 mm}
      \caption{Anomaly detection results on HDFS dataset}
      \label{fig:hdfs-anomaly-detection}
      \vspace{-5 mm}
\end{figure}

\begin{figure}[t]
      \begin{minipage}{1.0\linewidth}
      \centering
      \includegraphics[height=3cm,width = 6.5 cm]{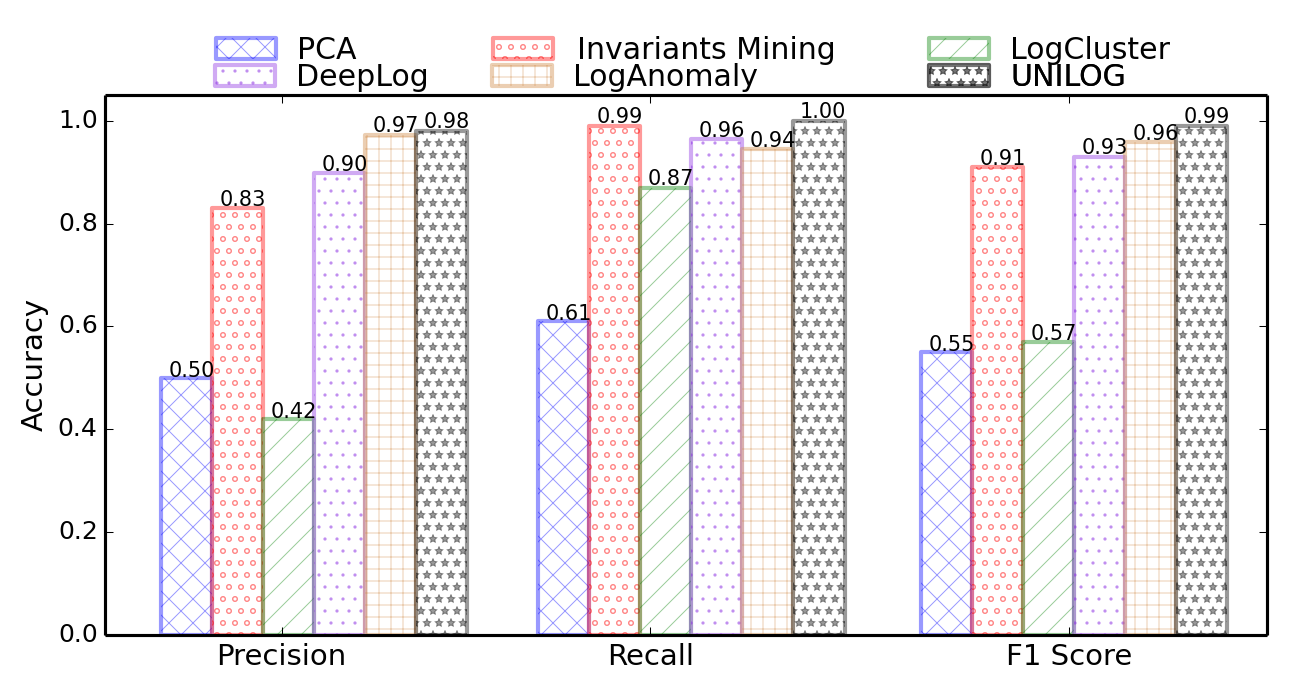}\\
      \end{minipage}
     \vspace{-3 mm}
      \caption{Anomaly detection results on BGL dataset.}
      \label{fig:bgl-anomaly-detection}
      \vspace{-5 mm}
\end{figure} 
\subsection{Setup}
\textbf{Implementation Details}
We follow the model design in Section 4. Specifically, we use 3 blocks for both in encoder and decoder. Each block consists of a multi-head self-attention layer, optional encoder-decoder attention layer, and a feed-forward network. We use 4 heads for self-attention, 32 inner dimensions for metrics in the attention layer, a hidden size of 128 for transformer blocks, and an output dimension of 512 for FFN. We apply Dropout with 0.3 probability in the model for regularization. Unlike This result modern network designed for NLP tasks, our model's size is four numbers of magnitude smaller since the scale of the public dataset for log analysis is much smaller than those for NLP tasks. This enables us to achieve good performance for log analysis tasks without the overwhelming number of parameters. 

All experiments are done on NVIDIA V100 GPUs 32 GB memory with Python 3.5 and Pytorch 1.3. We use NVIDIA Apex framework for distributed training. We use a batch size of 128, and trained with AdamW \cite{loshchilov2018fixing}. We use 180 as the maximum length of log sequence and pretrained with $2^16=65,536$ step. We set the initial learning rate as 5e-4 without learning rate warm-up stage and use exponential learning rate decays. We fine-tuned all tasks for $2^{11}=2048$ steps to prevent over-fitting.
\\
\\
\textbf{Datasets.} We use the BGL \cite{oliner2007supercomputers}, HDFS \cite{xu2009largescale}, HPC \cite{he2016experience}, ZooKeeper \cite{he2020loghub}, Proxifier \cite{zhu2019tools}, Hadoop \cite{lin2016log} and Switch \cite{zhang2018prefix} to conduct our experiments. We use the dataset split strategy from the prior works for fair comparison. For anomaly detection, failure prediction, summarization and compression we follows the Loganomaly~\cite{meng2019loganomaly}, PreFix~\cite{zhang2018prefix}, LogSummary~\cite{meng2020summarizing}, and LogParse~\cite{logzip}, respectively. For the pretrained dataset, We utilize all log datasets mentioned above to pretrain \name{}, except test data which we used for evaluation. 
\\
\\
\textbf{Evaluation Metrics.} For anomaly detection, summarization, and failure prediction, A method's capability is usually assessed by four metrics that have intuitive interpretation, \textit{i.e.,} precision, recall, and F1 score, and thus we use these metrics to evaluate each method. For log compression, we use compression rate $C=\frac{\textit{size of compressed logs}}{\textit{size of original logs}}$, which indicates the model's ability to store the same amount of logs with the number of storage space.

% \begin{figure}[t]
%       \begin{minipage}{1.0\linewidth}
%       \centering
%       \includegraphics[height=3cm, width = 6.5 cm]{./figures/hdfs_bar_long.png}\\
%       \end{minipage}
%      \vspace{-3 mm}
%       \caption{Anomaly detection results on HDFS dataset}
%       \label{fig:hdfs-anomaly-detection}
%       \vspace{-5 mm}
% \end{figure} 

% \begin{figure}[t]
%       \begin{minipage}{1.0\linewidth}
%       \centering
%       \includegraphics[height=3cm,width = 6.5 cm]{./figures/bgl_bar_long.png}\\
%       \end{minipage}
%      \vspace{-3 mm}
%       \caption{Anomaly detection results on BGL dataset.}
%       \label{fig:bgl-anomaly-detection}
%       \vspace{-5 mm}
% \end{figure} 

\begin{table}
\caption{Comparison of model performance for log compression task.}
\centering
% \small
% \renewcommand\tabcolsep{4.5 pt} % 调整表格列间的宽度
\begin{tabular}{c|cccc|c}
\toprule
Method &   HPC   &  HDFS   & Zoo   & Hadoop &    \textit{Avg.} \\
\midrule
bzip            & 5.7\%   & 9.3\%  &  3.0\% & 5.6\%  & 6.1\%  \\
7zIp            & 7.0\%   & 8.5\%  & 3.1\%  & 5.0\%  &5.9\%   \\
zip             & 8.6\%   & 11.4\% & 4.8\%  & 8.0\%  &8.2\%   \\
LogParse        & 23.4\%  & 14.1\% & 10.9\% & 42.4\% & 22.7\%  \\
LogParse + bzip & 3.4\%  & 3.2\%  & 2.4\% & 3.6\%  & 3.2\% \\
LogParse + zip  & 4.4\%  & 7.8\%  & 2.8\%  & 5.0\%  & 5.0\% \\
LogParse + 7zip      & 3.7\%   & 6.3\% &  2.8\%  & 3.4\%  & 4.1\%  \\
\textbf{\name{}}  & 2.7\%   & 2.9\% & 2.8\%  & 3.3\% &  2.9\%  \\
\bottomrule
\end{tabular}
\label{tbl:compress}
\end{table}
\subsection{Results.}
\textbf{Log Anomaly Detection.} Log anomaly detection is the most widely studied topic in the log analysis domain. In this section, we conduct experiments on two log anomaly detection benchmark: HDFS and BGL to demonstrate the effectiveness of our proposed framework. 
We compare \name{} to the following methods with 1) PCA~\cite{xu2009largescale}, PCA-based anomaly detection method vectorize each log sequence as an event count vector and employed to find patterns between the dimensions of event count vectors. 2) Invariant Mining~\cite{lou2010mining} which was the ﬁrst applied to log- based anomaly detection in \cite{lou2010mining} based on the assumption that program invariants are the linear relationships that always hold during system running. 3) LogCluster~\cite{lin2016log} that determine the state of a new log sequence by computing its distance to representative vectors in knowledge base and report the log sequence as normal or abnormal based on representation vector distance. 4) DeepLog~\cite{lin2016log} is the first deep learning approach on log anomaly detection. 5) LogAnomaly~\cite{meng2019loganomaly}, the state-of-the-art method on log anomaly detection, combines sequential patterns and quantitative patterns.
% \begin{itemize}
%     \item PCA~\cite{xu2009largescale}:The PCA-based anomaly detection method vectorize each log sequence as an event count vector and employed to find patterns between the dimensions of event count vectors.

%     \item Invariant Mining~\cite{lou2010mining}: Invariants mining was ﬁrst applied to log- based anomaly detection in \cite{lou2010mining} based on the assumption that program invariants are the linear relationships that always hold during system running.

%     \item LogCluster~\cite{lin2016log}:  LogCluster determine the state of a new log sequence by computing its distance to representative vectors in knowledge base and report the log sequence as normal or abnormal based on representation vector distance.
    
%     \item DeepLog~\cite{lin2016log}: Deeplog is the first deep learning approach on log anomaly detection, which utilizes session windows/fixed windows to get log sequences and utilizes LSTM to train a prediction model and learn normal patterns. 
%     \item LogAnomaly~\cite{meng2019loganomaly}: LogAnomaly is the state-of-the-art method on log anomaly detection, combines sequential patterns and quantitive patterns. It extracts semantic information of templates and uses template approximate to deal with the new type of logs in real-time. 
% \end{itemize}
In Fig.~\ref{fig:hdfs-anomaly-detection} and Fig.~\ref{fig:bgl-anomaly-detection} shows the experimental results for log anomaly detection on HDFS and BGL dataset. Our proposed \name{} achieves the best performance on both datasets. Specifically, on the HDFS dataset, \name{} achieved precision of 0.97, recall of 0.99, and f1 score of 0.98. The f1 score is 0.3\% higher than LogAnomaly. On the BGL dataset, our method achieve precision of 0.98, recall of 1.00 and f1 score of 0.99. It further improve the f1 score 0.3\% over LogAnomaly. This improvement is non-trivial, since reliability on the system is vital and an f1 score improve over 0.1\% might save millions of dollars.
\\
\\
\begin{table}
\caption{Comparison of model performance for log summarization task.}
\centering
% \small
% \renewcommand\tabcolsep{4.5 pt} % 调整表格列间的宽度
\begin{tabular}{c|cccc}
\toprule
             Logs &      Method& Precision & Recall & F1  \\
\midrule
                  
\multirow{3}*{BGL}
%&   LDA      &   0.382   &  0.076 &   0.119 \\
                  & TextRank   &   0.893   &  0.238 &   0.347 \\
 %                 & TF-IDF    &   0.383   &  0.354 &   0.332 \\
~& LogSummary & 0.815 & 0.703   & 0.788 \\

&  \textbf{\name{}} & 0.895 & 0.824 & 0.858   \\
                  
\midrule
\multirow{3}*{HDFS}

%&   LDA      &   0.382   &  0.076 &   0.119 \\
                  & TextRank   &   0.893   &  0.238 &   0.347 \\
  %                & TF-IDF    &   0.383   &  0.354 &   0.332 \\

~& LogSummary &0.759 &0.432& 0.538\\
&  \textbf{\name{}} & 0.876 &0.650 & 0.746\\

\midrule
\multirow{3}*{HPC} 
%&LDA      &   0.530   &  0.110 &   0.175 \\
& TextRank   &   0.904   &  0.265 &   0.365 \\
%&  TF-IDF    &   0.487   &  0.506 &   0.472 \\
& LogSummary&0.819&0.911&0.840\\
& \textbf{\name{}}&0.835&0.911&0.871 \\

\midrule 
\multirow{3}*{Proxifier}   
%&LDA      &   0.332   &  0.088 &   0.135 \\
& TextRank   &   0.663   &  0.050 &   0.093 \\
%& TF-IDF&   0.281   &  0.324 &   0.290 \\
&   LogSummary &0.879&0.857 &0.864 \\                  
& \textbf{\name{}} &0.925&0.877 &0.900 \\
\bottomrule
\end{tabular}
\vspace{-3mm}
	\label{tbl:summary}
\end{table}
\begin{table}
\caption{Comparison of compression rate for Log data compression task with Post-LN versus Pre-LN structure.}
\centering
% \small
% \renewcommand\tabcolsep{4.5 pt} % 调整表格列间的宽度
\begin{tabular}{cccccc}
\toprule
Task & Dataset &  Method   &  Precision   & Recall  & F1 \\
\midrule
\multirow{2}{*}{A.D.} &\multirow{2}{*}{BGL}& Post-LN &  0.95 & 0.94 &  0.94\\
& & Pre-LN & 0.98  & 1.00  & 0.99 \\
\midrule
\multirow{2}{*}{Summ.} & \multirow{2}{*}{BGL} & Post-LN  &  0.850 & 0.832 & 0.841 \\
& & Pre-LN  &  0.895  & 0.824 & 0.858 \\
\bottomrule
\end{tabular}
\label{tbl:preln}
\end{table}

\textbf{Log Summarization.} After demonstrating the superiority of our method on log anomaly detection, we evaluate it for log summarization that requires refining the semantic information in the log sequence.   We demonstrate the experimental results for log data summarization. We choose TextRank \cite{textrank}, a classic keyword extraction algorithm, and LogSummary \cite{meng2020summarizing} the state-of-the-art method design specifically to summarize log data. Table~\ref{tbl:summary} reports the experimental results; it shows that the \name{} can significantly improve our log summarization performance: \name{} is improving the F1 score by 0.070 on BGL data, 0.031 on HPC data, and 0.036 on Proxifier over LogSummary. On hard task HDFS, the F1 score of \name{} is 0.208 higher than LogSummary, demonstrate that \name{} is more effective than the classic model and LogSummary in a real-world situation.
\\
\\
\noindent
\textbf{Log Failure Prediction.} We now show \name{} can improve the performance on failure prediction tasks. As shown in Table~\ref{tbl:fail}, we perform experiment on Switch log datasets to compare \name{} against a) SKSVM \cite{fulp2008predicting}, a log-based failure prediction system used in computer b) HSMM \cite{salfner2007using}, a log-based failure prediction in ISP devices, and c) Prefix \cite{zhang2018prefix}, the state-of-the-art log failure prediction algorithm for switch failure prediction. We can observe that, though both SKSVM and HSMM are failure prediction algorithms, their performance is relatively low. This is due to that these two methods are designed specifically for computer and ISP service. As a result, they fail to generalize on Switch log dataset. On the contrary, our \name{} achieves better performance on all three switches, despite the fact that \name{} is not being designed for switch failure prediction tasks. Particularly, on the M2 and M3 switch, the \name{} strongly outperforms these approaches notably, obtains a +5.27\% and +9.85\% F1 score improvement. This provides another indication that \name{} is able to capture generalizable knowledge in the cross-service log data.
\\
\\
\noindent
\textbf{Log Compression.} We also conduct experiments on log compression. We compare our method to three popular data compression techniques that are widely used in the computer system, bzip \cite{bzip}, 7zip \cite{7zip}, zip \cite{zip}. We also compare to LogParse~\cite{meng2020logparse}, a log parsing method that can adaptively classify log based on the word. The LogParse can effectively combine with existing data compression techniques to compress log data. We therefore compare our method to LogParse, which is followed by bzip, 7zip, and zip algorithm. The experimental results are shown in Table~\ref{tbl:compress}. Our method consistently improves the compression ratio over all baseline methods. Specifically, \name{} increase 2.1\% on average compression rate on four datasets compare to the previous state-of-the-art, which results in an excellent storage reduction. Besides, as the number of samples in the target dataset increase, \name{} achieve more improvement on the performance. For instance, on the largest dataset HDFS, where the number of logs is over two orders of magnitudes larger than HPC and Hadoop, the \name{} improves the compression rate relatively up to 3.4\%.

%One significant advantage of a learnable algorithm is that the effectiveness increase as the number of samples in the target dataset increase. This observation is verified by the experimental results. Specifically, on a small dataset such as ZooKeeper, the compression rate of \name{} is equivalent to LogParse, a rule-based method. On larger datasets such as HPC, HDFS, and Hadoop, The boost on compression rate is significant. On HPC, the \name{} achieves a 2.7\% compression rate, which is 1\% higher than LogParse*, and on Hadoop, \name{} improves 0.1\% over LogParse. On the largest dataset HDFS, where the number of logs is over two orders of magnitudes larger than HPC and Hadoop, the \name{} improves the compression rate relatively up to 3.4\%. This is achieved by extracting better data-specific semantic information. As more common rules in the dataset are captured by the algorithm, more pieces of data can be compressed without loss of information. The \name{} shows the great potential of our model on the real-world log data compression application.
\begin{table}[t]
\caption{Comparison of model performance for Log data compression task with different activation layer.}
\centering
% \small
% \renewcommand\tabcolsep{4.5 pt} % 调整表格列间的宽度
\begin{tabular}{lccccc}
\toprule
Task & Dataset &  Method   &  Precision   & Recall  & F1 \\
\midrule
\multirow{2}{*}{A.D.} &\multirow{2}{*}{BGL}& ReLU &  0.97 & 0.99 &  0.98\\
& & LogAct & 0.98  & 1.00  & 0.99 \\
\midrule
\multirow{2}{*}{Summ.} & \multirow{2}{*}{BGL} & ReLU  &  0.866 & 0.815 & 0.840 \\
& & LogAct  &  0.895  & 0.824 & 0.858 \\
\bottomrule
\end{tabular}
\label{tbl:relu}
\end{table}
\begin{table}[t]
\caption{Comparison of model performance for Log data compression task with different pretrain objectives.}
\centering
% \small
\renewcommand\tabcolsep{3.5 pt} % 调整表格列间的宽度
\begin{tabular}{cccccc}
\toprule
Task & Dataset &  Method   &  Precision   & Recall  & F1 \\
\midrule
\multirow{3}{*}{A.D.} & \multirow{3}{*}{BGL} & 
Prefix Language Model &  0.65 & 0.68 &  0.66\\
& & BERT-style &  0.95 & 0.94 &  0.94\\
& & \textbf{\name{}} & 0.98  & 1.00  & 0.99 \\
\midrule
\multirow{3}{*}{Summ.} & \multirow{3}{*}{HDFS} & 
Prefix Language Model &  0.325 & 0.739 &  0.451\\
& & BERT-style &  0.770 & 0.521 & 0.621\\
& & \textbf{\name{}} & 0.876  & 0.650  & 0.746 \\
\bottomrule
\end{tabular}
\label{tbl:pretrain_comparison}
\end{table}

\subsection{Ablation Study}

\input{ablation}

\label{sec:ablation}

%% file: ablation.tex
\begin{table}[t]
\caption{Comparison of model performance for Log data compression task with different pretrain objectives.}
\centering
% \small
\renewcommand\tabcolsep{3.5 pt} % 调整表格列间的宽度
\begin{tabular}{cccccc}
\toprule
Task & Dataset &  Method   &  Precision   & Recall  & F1 \\
\midrule
\multirow{3}{*}{A.D.} & \multirow{3}{*}{BGL} & 
Prefix Language Model &  0.65 & 0.68 &  0.66\\
& & BERT-style &  0.95 & 0.94 &  0.94\\
& & \textbf{\name{}} & 0.98  & 1.00  & 0.99 \\
\midrule
\multirow{3}{*}{Summ.} & \multirow{3}{*}{HDFS} & 
Prefix Language Model &  0.325 & 0.739 &  0.451\\
& & BERT-style &  0.770 & 0.521 & 0.621\\
& & \textbf{\name{}} & 0.876  & 0.650  & 0.746 \\
\bottomrule
\end{tabular}
\label{tbl:pretrain_comparison}
\end{table}

\begin{table}[t]
\caption{Comparison of model performance on log anomaly detection, log compression, and log summarization on HDFS data.}
\centering
% \small
\renewcommand\tabcolsep{3.5 pt} % 调整表格列间的宽度
\begin{tabular}{c|c|ccc|c}
\toprule
Pretrain Dataset &  Task  &  Precision   &  Recall   & F1  & Comp. Rate\\
\midrule
\multirow{3}{*}{HDFS}  &  A.D. & 0.94  & 0.96  & 0.95  &  - \\
&    Comp. &  - &  - & -  & 3.5\% \\
&    Summ. & 0.807  & 0.545  & 0.651  & - \\
\midrule
\multirow{3}{*}{+ BGL} &  A.D. &  0.90 & 0.93  & 0.91  & - \\
&   Comp. &  - & -  & -  & 5.6\% \\
&    Summ. & 0.652  & 0.380  & 0.480  & - \\
\midrule
% \multirow{3}{*}{+ Proxifier + ZooKeeper + Hadoop + HPC} &  A.D. &  0.94 & 0.94  &  0.94 & -  \\
+ (Proxifier  &  A.D. &  0.94 & 0.94  &  0.94 & -  \\
+ ZooKeeper &    Comp. & -  & -  & -  & 5.6\% \\
+ Hadoop + HPC)&    Summ. &  0.812 &  0.597 & 0.688  & - \\
\midrule
\multirow{3}{*}{+ Switch}  &  A.D. & 0.97  &  0.99 & 0.98  & - \\
&    Comp. &  - & -  & -  & 2.9\% \\
&    Summ. & 0.876  &  0.650 & 0.746  & - \\
\bottomrule
\end{tabular}
\label{tbl:more_data_good}
\vspace{-2 mm}
\end{table}

\begin{table}[t]
\caption{Comparison of model performance for transfer learning to Hadoop.}
\centering
% \small
% \renewcommand\tabcolsep{4.5 pt} % 调整表格列间的宽度
\begin{tabular}{c|ccc}
\toprule
Method   &  Precision   & Recall  & F1 \\
\midrule
LogTransfer & 0.980  &  0.970 & 0.977 \\
\textbf{\name{}}  & 1.000  & 0.986  & 0.993 \\
\bottomrule
\end{tabular}
\label{tbl:transfer_learning}
\vspace{- 2mm}
\end{table}

\textbf{Pretrain Objectives.} We conduct experiments to validate the effectiveness of denoising objectives in our method. We compare with 1) Prefix Language Modeling, where log sequences are feed into the network and ask to return the next log sequence, 2) the BERT-style objective, where consecutive tokens are masked. In Table~\ref{tbl:pretrain_comparison}, we present the experimental results on BGL anomaly detection task and HDFS summarization task. The prefix language modeling is only obtained an F1 score of 0.66, which is not suitable for the log analysis task. On both tasks, the \name{} with our proposed pretrain objective achieve significantly better results than both prefix language modeling and BERT-style objective. On log summarization, which is a semantic essential task, our pretrain objective outperform BERT-style objective by 0.125 on F1 score, shows that our method is more suitable to extract representation in the unstructured log data.
\\
\\
\textbf{Pre-LN.} We evaluate the importance of Pre-LN architecture in Table~\ref{tbl:preln}. The experiments are conducted on BGL dataset with log anomaly detection and log summarization. We can observe that the experimental results on both tasks are improved. On the log summarization tasks, the F1 score of the model using LogAct is 0.018\% higher than the model using ReLU. The improvement on precision is significant, which is over 0.031\%.  
\\
\\
\textbf{LogAct.}
We verfy the effectiveness of our proposed LogAct layer. We evaluate the model performance on the BGL dataset on log anomaly detection and log summarization tasks, and the result is shown in Table~\ref{tbl:relu}. The precision, recall, and F1 score on Summarization of LogAct is lifed by, respectively. On the other hand, the On log anomaly detection, the LogAct increases on all evaluation metrics by 1\%. Because the performance on BGL dataset is already high, thus these improvement is non-trivial. The results indicate the complex feature representation that unstructured text may have and, therefore, can benefit from more non-linearity.
\\
\\
\textbf{Does Transformer Really Benefit from More Unstructured Log Data?} The log data is considered to be data-intensive yet lacks useful information. Hence we study that if massive log data can really achieve better performance than small-scale log data. We evaluate our assumption by progressively adding more datasets into our pretrain model and then finetune on the task model. We linearly reduce the total training iterations based on the number of log data in the total pretrain dataset. The rest of the training strategy is kept the same. We evaluate on HDFS to validate our assumption. We provide the results on three tasks, log anomaly detection, log compression, and log summarization. The Table~\ref{tbl:more_data_good} shows the experimental results. We can observe that when the HDFS is the only pretrain dataset, the result on sub-tasks is relatively higher than using HDFS + BGL + Proxifier + ZooKeeper. This is intuitive since directly pretrain on the target data can certainly improve the performance on finetuning. This phenomenon is utilized in the self-training~\cite{chen2020simple}. When the number of pretrain data is not large enough, the model's eventual performance on target data may be bias towards other datasets. However, the performance is improved and eventually surpasses the performance on sole pretrained on HDFS. Especially when we add the Switch dataset, the F1 score on anomaly detection increases from 0.95 to 0.98, the compression rate increases from 3.5\% to 2.9\%, and the F1 score on summarization is achieved 0.651 compared to 0.746. This supports our assumption that the \name{} can indeed improve by training on more log data, though the information entropy provided by the log data is small.
\\
\\
\textbf{Does \name{} generalize on unseen dataset?} As we have previously stated that, \name{} can learn common features from massive log data via pretrain. In the real world, the algorithm running on the server can often feed with unseen data, which results in poor performance due to the \textit{data domain shift}. Though our method does not target to solve domain adaptation problems, we evaluate the ability of \name{} to generalize on unseen data. We pretrain our model on our pretrain dataset without the Hadoop dataset. Then, we evaluate our method on the Hadoop dataset by finding anomalous samples in the Hadoop. We compare our method with LogTransfer~\cite{chen2020logtransfer}, a log-based domain adaptation method. The LogTransfer is trained on the HDFS dataset and then evaluate on the Hadoop dataset. Table~\ref{tbl:transfer_learning} shows that, with the pretrain-finetune paradigm, the transfer learning ability of \name{} is better than LogTransfer, outperforms the state-of-the-art method by 0.16 on the F1 score.

%% file: conclusion.tex
This work aim to build a unified framework for log analysis tasks. We propose \textit{UniLog}, which enable the engineers to deploy one model that are specialized for all log analysis tasks. Our experiments have shown that UniLog perform competitively with, or outperform, SOTA methods on various log datasets across four log analysis tasks.